\documentclass[aps,prl,twocolumn,superscriptaddress,showpacs]{revtex4}
\usepackage{graphicx}

\begin{document}


\title{Spin dynamics and magnetic order in magnetically frustrated
Tb$_2$Sn$_2$O$_7$}



\author{P.~Dalmas de R\'eotier}
\affiliation{CEA/DSM/D\'epartement de Recherche
Fondamentale sur la Mati\`ere Condens\'ee, 38054 Grenoble, France}

\author {A.~Yaouanc}
\affiliation{CEA/DSM/D\'epartement de Recherche
Fondamentale sur la Mati\`ere Condens\'ee, 38054 Grenoble, France}

\author{L.~Keller}
\affiliation{Laboratory for Neutron Scattering, 
ETH Z\"urich and Paul Scherrer Institute, 5232 Villigen-PSI, Switzerland}

\author{A.~Cervellino}
\altaffiliation[On leave from]{
CNR, Istituto di Cristallografia (CNR-IC), 70126 Bari, Italy.}
\affiliation{Laboratory for Neutron Scattering, 
ETH Z\"urich and Paul Scherrer Institute, 5232 Villigen-PSI, Switzerland}

\author{B.~Roessli}
\affiliation{Laboratory for Neutron Scattering, 
ETH Z\"urich and Paul Scherrer Institute, 5232 Villigen-PSI, Switzerland}

\author{C.~Baines}
\affiliation{Low temperature facilities group, 
Paul Scherrer Institute, 5232 Villigen-PSI, Switzerland}

\author{A.~Forget}
\affiliation{CEA/DSM/D\'epartement de Recherche
sur l'Etat Condens\'ee, les Atomes et les Mol\'ecules, 
91191 Gif sur Yvette, France}

\author{C.~Vaju}
\affiliation{CEA/DSM/D\'epartement de Recherche
Fondamentale sur la Mati\`ere Condens\'ee, 38054 Grenoble, France}

\author{P.C.M.~Gubbens}
\affiliation{Department of Radiation, Radionuclides \& Reactors,
Delft University of Technology, 2629 JB Delft, 
The Netherlands}

\author{A.~Amato}
\affiliation{Laboratory for Muon-Spin Spectroscopy, 
Paul Scherrer Institute, 5232 Villigen-PSI, Switzerland}

\author{P.J.C.~King}
\affiliation{ISIS Facility, Rutherford Appleton Laboratory, Chilton, Didcot, 
OX11 0QX, UK}



\date{\today}

\begin{abstract}

We report a study of the geometrically frustrated magnetic material Tb$_2$Sn$_2$O$_7$ by the positive muon spin relaxation technique. 
No signature of a static magnetically ordered state is detected while neutron magnetic reflections are observed in agreement with a published report. 
This is explained by the dynamical nature of the ground state of 
Tb$_2$Sn$_2$O$_7$: the Tb$^{3+}$ magnetic moment characteristic fluctuation 
time is $\simeq$ $10^{-10}$ s. The strong effect of the magnetic field on the muon 
spin-lattice relaxation rate at low fields indicates a large field-induced
increase of the 
magnetic density of states of the collective excitations at low energy.

\end{abstract}

\pacs{75.40.-s, 75.25.+z, 76.75.+i}

\maketitle

Magnetic materials with antiferromagnetically coupled spins located on triangular motifs 
exhibit geometrical magnetic frustration
because their spatial arrangement is such that it prevents the simultaneous
minimization of all the interaction energies \cite{Villain79}. The frustration, which leads 
to a highly degenerate ground state, forbids magnetic order to occur. Perturbations to the nearest-neighbor 
exchange interaction, such as exchange interactions extending beyond nearest-neighbor 
magnetic atoms, dipole coupling or magnetic anisotropy, are believed to be responsible for the magnetic order
observed in some compounds \cite{Ramirez01}. Typical examples are given by the spinel structure oxide LiMn$_2$O$_4$,
the pyrochlore structure compounds Gd$_2$Ti$_2$O$_7$, Er$_2$Ti$_2$O$_7$, Tb$_2$Sn$_2$O$_7$ and Gd$_2$Sn$_2$O$_7$ and
the cuprate mineral Cu$_2$Cl(OH)$_3$. According to neutron diffraction (ND) measurements, long- and short-range orders coexist 
in LiMn$_2$O$_4$ \cite{Greedan02}, a partial order of the Gd$^{3+}$ magnetic moments is established at low temperature in 
Gd$_2$Ti$_2$O$_7$ \cite{Stewart04} and a conventional magnetic order exists for the last three pyrochlore compounds
\cite{Champion03,Mirebeau05,Wills06}. When looked for, persistent spin dynamics has always been found far below the magnetic ordering
temperature. The possibility of such dynamics is conceivable for the partly ordered structure of 
Gd$_2$Ti$_2$O$_7$ \cite{Yaouanc05} and Cu$_2$Cl(OH)$_3$ \cite{Zheng05} but more exotic when all the magnetic 
moments contribute to the magnetic structure as for 
Er$_2$Ti$_2$O$_7$ \cite{Lago05}, Gd$_2$Sn$_2$O$_7$ \cite{Bertin02,Bonville04a} 
and Tb$_2$Sn$_2$O$_7$ \cite{Mirebeau05}. A prerequisite for
understanding the unanticipated behavior of these latter systems is
a careful characterization of their dynamical properties.

Here we show that positive muon spin relaxation ($\mu$SR) 
and ND results in the ordered phase 
of Tb$_2$Sn$_2$O$_7$ can be simultaneously accounted 
for only if the Tb$^{3+}$ moments are strongly dynamical. An independent
and consistent time scale is obtained from a careful analysis of the
neutron data. In addition, the initial strong and 
counter-intuitive increase of the muon relaxation rate when a magnetic field
is applied indicates an increase of the density of magnetic excitations at very
low energy.

Tb$_2$Sn$_2$O$_7$ crystallizes with the cubic space group $Fd{\bar 3}m$. Rietveld refinements of powder x-ray and ND patterns 
yield the lattice constant $a$ = 10.426~{\AA} and the free position parameter allowed by the space group for the $48f$ site occupied by oxygen, $x= 0.336$ \cite{Mirebeau05}. Magnetic measurements
point to a magnetic transition at 0.87 K and to strong antiferromagnetic interactions as deduced from the large and negative Curie-Weiss 
constant $\theta_{\rm CW} = -12\, {\rm K}$  \cite{Matsuhira02}. Powder ND indicates a structure with both ferromagnetic 
and antiferromagnetic components below $T_{\rm sr}$ = 1.3 (1) K where short-range magnetic correlations which are not liquid-like
appear \cite{Mirebeau05}. A steep increase of the Tb$^{3+}$ magnetic moment $\mu_{\rm Tb}$ and correlation length $L_c$ is observed around $T_{\rm lr}$ = 
0.87~K where a peak is seen in the temperature dependence of the 
specific heat $C_p(T)$. 
$\mu_{\rm Tb} = 5.9 \,(1)\, \mu_{\rm B}$ with $L_c$ = 19 nm at 
0.10 K. Therefore, even far below $T_{\rm lr}$, $L_c $ is much shorter than
usually observed in ordered magnetic structures.

We present below (i) $C_p(T)$ data recorded using a dynamic 
adiabatic technique, (ii) ND measurements carried out at the cold neutron 
powder diffractometer DMC of the SINQ facility at the Paul Scherrer 
Institute and (iii) $\mu$SR measurements done at the MuSR spectrometer of 
the ISIS facility (Rutherford Appleton 
Laboratory, Chilton, United Kingdom) and GPS and LTF spectrometers 
of the Swiss Muon Source (Paul Scherrer Institute, Villigen, Switzerland).

\begin{figure}
\includegraphics[scale=0.8]{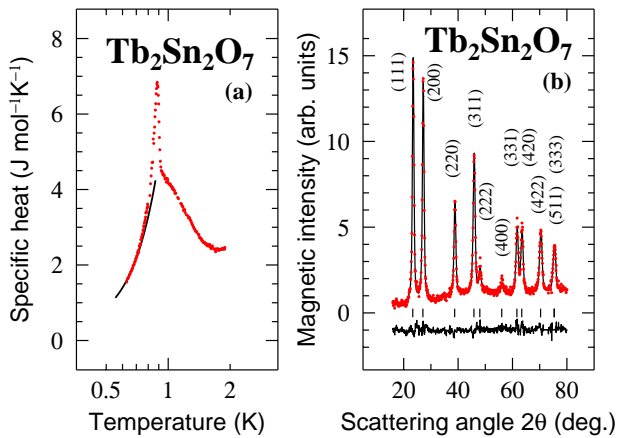}
\caption{(color online).
(a). Low temperature dependence of the specific heat per mole of 
Tb measured for our Tb$_2$Sn$_2$O$_7$ sample. The sharp maximum is the signature of a magnetic transition occurring at $T_{\rm lr}$ = 0.88 K. The solid line
is a fit to a model explained in the main text.
(b). Magnetic powder diffraction pattern of Tb$_2$Sn$_2$O$_7$ versus the scattering angle 
$2 \theta$ obtained from the subtraction of data recorded at 0.11~K and 1.23~K. Neutrons of wavelength $2.453$ {\AA} were used. The solid 
lines show the best refinement and the difference spectrum (bottom).  From the pattern analysis we extract
a Tb$^{3+}$ magnetic moment $\mu_{\rm Tb} = 5.4 \,(1)\, \mu_{\rm B}$ which makes an angle of 14.2$^\circ$ with the local $\langle 111 \rangle$
axis. A measurement (not shown) at 0.98 K gives  $\mu_{\rm Tb} = 1.6 \,(2) \, 
\mu_{\rm B}$ and a Rietveld refinement of the diffraction pattern recorded at 
100 K is consistent with space group $Fd{\bar 3}m$ (lattice and oxygen
position parameters equal to $a$ = 10.427~{\AA} and $x$ = 0.337). }
\label{Sh_neutron}
\end{figure}

We first present $C_p(T)$ and powder ND measurements;
see Fig.~\ref{Sh_neutron}. Relative to the published 
$C_p(T)$ data \cite{Mirebeau05}, our sample displays a peak somewhat 
stronger in intensity at $T_{\rm lr} = 0.88$  K.  
The magnetic ND pattern at low temperature is in very reasonable agreement with 
published results \cite{Mirebeau05}.
The magnetic reflections are not resolution limited and their tails are 
Lorentzian-like rather than the usual Gaussian-like. Their shape has been
fitted in detail assuming a size distribution of sharply defined spherical
magnetic domains. A magnetic reflection is then represented as the 
distribution-weighted sum of the peak profiles of domains of different sizes,
each of them being convoluted with the instrument resolution function
\cite{Cervellino05}. Two types of
distributions, namely the log-normal and gamma distributions, were used
for the fit of the data shown in Fig.~\ref{Sh_neutron}b and gave similar 
results. For instance the average radii for
the domains are 3.14 (10) and 2.95 (15) nm 
respectively for the two distributions. The volume-averaged domain diameter 
$D_v$ $\equiv$ $L_c$ can
be deduced from the third and fourth moments $M_3$ and $M_4$ of the 
distribution using $D_v$ = $3M_4/(2M_3)$ \cite{Popa02}. 
Numerically, we find 19.8 and 19.1 nm for each of the considered distributions
in excellent agreement with the value of $L_c$ given above.

\begin{figure}
\includegraphics[scale=0.8]{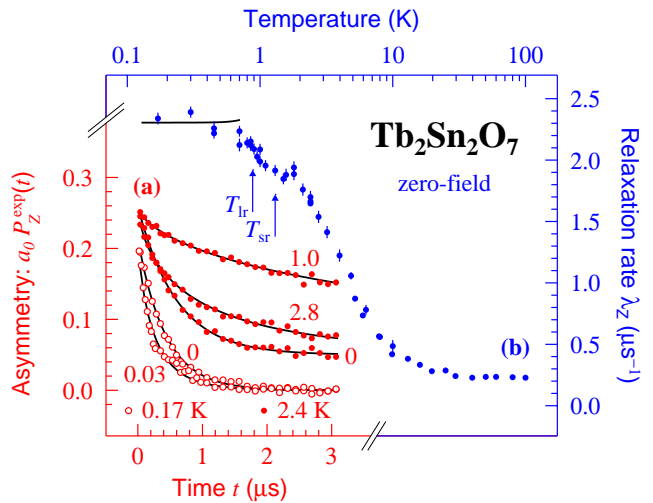}
\caption{(color online). (a).
Two sets of $\mu$SR spectra recorded respectively at 0.17 and 2.4 K. 
For clarity the latter ones have been vertically shifted by 0.05.
The numbers written next to each spectrum correspond to the longitudinal
magnetic field, expressed in tesla, for which the data have been recorded.
The solid lines are fits to the stretched exponential relaxation
function, $P_Z^{\rm exp}(t) = \exp[(-\lambda_Z t)^{\alpha}]$, where $\alpha$ 
= 1
in zero-field. 
(b). Spin-lattice relaxation rate $\lambda_Z$ 
versus temperature, deduced from spectra recorded using a cooling down 
sequence. The data were taken in zero-field, with the exception
of some spectra above 4 K which were recorded in a longitudinal field 
$B_{\rm ext}$ = 2 mT. All the corresponding spectra were fitted 
with $\alpha$ = 1.
The solid
line is deduced from the density of states introduced in the main text. 
}
\label{muon_zf}
\end{figure}

Now we report on the $\mu$SR data: 
see Refs.~\cite{Dalmas97,Dalmas04} for an introduction to this technique. 
In  Fig.~\ref{muon_zf}a two zero-field spectra are presented, one recorded at 
a temperature $T > T_{\rm sr}$ and 
a second deep in the ordered state {\sl i.e.} $T \ll T_{\rm lr}$.
Unexpectedly, the two spectra are 
qualitatively similar, {\sl i.e.} no clear-cut signature of the magnetic transition 
below $T_{\rm lr}$ is detected. We shall argue that this reflects the non-static character 
of the magnetic ground state of Tb$_2$Sn$_2$O$_7$. 

We recall that the zero-field $\mu$SR technique gives access to the longitudinal polarization 
function $P^{\rm exp}_Z(t)$ and, in the magnetically ordered state of a powder sample, 
it is expected to be the weighted sum of the longitudinal and transverse components: 
$P^{\rm exp}_Z(t) = [\exp(-\lambda_Z t) + 2 \exp(-\lambda_X t) \cos(\gamma_\mu \langle B_{\rm loc} \rangle t)]/3$. 
$\lambda_Z$ and $\lambda_X$ are, respectively, the spin-lattice and spin-spin relaxation rates,
$\gamma_\mu$ is the muon gyromagnetic ratio ($\gamma_\mu$ = 851.615 Mrad s$^{-1}$ T$^{-1}$) and 
$\langle B_{\rm loc} \rangle$ stands for the mean value of the local field at the muon site. Since
all the measured spectra are exponential-like, this requires $\lambda_X \simeq \lambda_Z$ and the oscillatory behavior to be absent. 

The first requirement is expected to be satisfied in the motional narrowing limit, {\sl i.e.}\ when the dynamics of the Tb$^{3+}$ moments is sufficiently 
fast; see for example Ref.~[\onlinecite{Slichter96}].
The disappearance of the oscillations in the transverse component can have two origins: either 
$\langle B_{\rm loc} \rangle =0$ or $\tau_c \ll (\gamma_\mu B_{\rm loc})^{-1}$
 where $\tau_c$ is the characteristic fluctuation time of $B_{\rm loc}$. 
We consider now these two possibilities.

The local field is built up from the dipole fields generated by the Tb$^{3+}$ magnetic moments. Taking into account the 
magnetic structure, we have mapped the dipole field in the unit cell. The 
computed field is small enough to be consistent with the experimental result 
only in the 
neighborhood of site (0.212, 0.537, 0.463) and symmetry equivalent sites. This 
site is located at $\simeq$ 1.4 {\AA} from the closest oxygen atom
neighbor. This corresponds to a much larger distance than the one
usually adopted by the muon site in oxides (range  1.0-1.1 {\AA}) (see e.g.
Refs.~[\onlinecite{Brewer90,Hitti90}]). 
Therefore this possibility is considered very unlikely. 

Since the hypothesis $\langle B_{\rm loc} \rangle =0$ does not hold we now
estimate a value of $\tau_c$ from our data. 
We assume for simplicity a transverse stochastic field jumping between two opposite orientations \cite{Kehr78}.
Generalization of this model would not change qualitatively the result.
Referring to the spontaneous fields measured for
Gd$_2$Ti$_2$O$_7$ \cite{Yaouanc05} and Gd$_2$Sn$_2$O$_7$ \cite{Bonville04a}, 
$B_{\rm loc}$ is estimated to be 0.2 T. Since with our model $\lambda_Z = \gamma_\mu^2 B_{\rm loc}^2 \tau_c$,
we compute $\tau_c = 8 \times 10^{-11}$ s from the measurements of 
$\lambda_Z$ at low temperature (Fig.~\ref{muon_zf}b). The whole $\mu$SR 
analysis is consistent since the motional narrowing condition is fulfilled 
($\gamma_\mu B_{\rm loc} \tau_c  \simeq 0.01 \ll 1$).

Now we have to understand the observation of magnetic reflections in neutron
scattering. The fact that these reflections can be indexed in the 
crystallographic structure of Tb$_2$Sn$_2$O$_7$ implies that the scattering 
is elastic or nearly elastic. 
The energy resolution of the DMC diffractometer given by the energy
spread of the incident neutrons is 
$\Delta E = 0.4$ meV, a typical value for this kind 
of instrument. In other words, such a scattering experiment probes the magnetic
structure with a time scale $\Delta t = \hbar /\Delta E= 1.6 
\times 10^{-12}$~s. Since $\tau_c \gg \Delta t$ 
the ND and $\mu$SR results are compatible. 
Now, a time scale directly related to our diffraction measurements can be 
estimated.
We use the relation $E=\hbar^2k^2/(2m_n)$ defining the energy of a 
neutron of wavevector $k$ and mass $m_n$. From a momentum width of 
0.04 \AA$^{-1}$ deduced from the average in ${\bf k}$ space of the domain size 
distribution, a time scale of at most $2 \times 10^{-10}$~s is obtained. 
It is rewarding that this value is of the same order of magnitude 
as $\tau_c$.

Our result implies that the neutron scattering is not purely elastic, 
{\sl i.e.} it occurs with a finite energy transfer. 
This dynamical order is consistent with the apparent reduction of the 
Tb$^{3+}$ moment deduced from the nuclear
specific heat relative to the neutron determination \cite{Mirebeau05}. 

$\lambda_Z(T)$ is presented in Fig.~\ref{muon_zf}b. The rate is first temperature independent from 
100 K down to $\sim$ 20 K and then starts to increase. This is consistent with the building up 
of pair-correlations at low temperatures \cite{Dalmas04}. $\lambda_Z$ increases steadily as the sample 
is cooled. Remarkably, no sharp anomalies are detected at either $T_{\rm sr}$ or $T_{\rm lr}$. 
$\lambda_Z (T)$ presents an inflection point around $T_{\rm lr}$. $T_{\rm sr}$ seems to
correspond to a change in the slope of $\lambda_Z(T)$. Below 0.7~K $\lambda_Z$ is
observed to be only weakly temperature dependent down to the lowest measured 
temperature.
That thermal behavior will be further 
discussed below.

\begin{figure}
\includegraphics[scale=0.8]{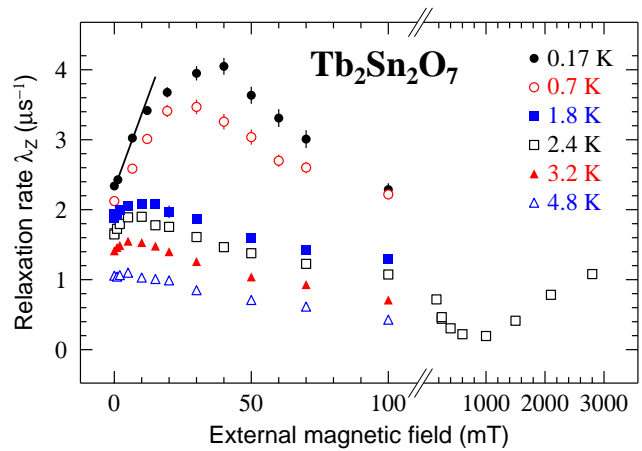}
\caption{(color online).
Field dependence of $\lambda_Z$ measured at low fields 
for Tb$_2$Sn$_2$O$_7$ using an experimental protocol explained in the main text.
An amazing initial increase of $\lambda_Z$ is observed at 3.2 K, 
{\sl i.e.}\ far above $T_{\rm sr}$, 
and down to 0.17 K. The shooting up of $\lambda_Z$ at high field displayed 
for $T = 2.4$ K is also quite unexpected.
The full line specifies the initial slope at 0.17 K which is 0.11 $\mu$s$^{-1}$mT$^{-1}$.}
\label{muon_lf}
\end{figure}

We have also recorded $\mu$SR spectra under longitudinal fields (see 
Fig~\ref{muon_zf}a). Whereas the sequence of the measurements 
has no importance for $T> T_{\rm sr}$, it matters for $T< T_{\rm sr}$, suggesting some hysteresis. 
In the latter case, the present results were deduced from spectra obtained by first cooling the sample 
in zero field from $T> T_{\rm sr}$ and then using a field-increase sequence for the recording.  
In contrast to the zero-field spectra, a stretched exponential function (see caption of Fig.~\ref{muon_zf})
with $\alpha <$ 1 is needed to fit these ones \cite{comments_Tb}. 
$\alpha$ (equal to 1 in zero-field) decreases smoothly with field to reach
a value of respectively $\sim 0.8$ for 
$T> T_{\rm sr}$ and $\sim 0.7$ for $T< T_{\rm sr}$ at 40 mT and then keeps this value up to 100 mT. 
$\lambda_Z(B_{\rm ext})$ is presented in Fig.~\ref{muon_lf}. The observed initial increase at low fields 
is unexpected. The slope is large and of approximately the same value for the two temperatures probed
below $T_{\rm lr}$ and is still detectable in the paramagnetic phase up to 3.2 K. Since 
$\lambda_Z$ decreases at large fields for all the temperatures, an extremum takes place at intermediate 
fields for $T \le 3.2$ K. Its position is located around 5 mT at 3.2 K and goes up
smoothly to reach a value $\sim$ 40 mT at 0.17 K, {\sl i.e.} almost an increase of an order of magnitude for more than a decade of temperature variation.  
The extremum being observed both in the ordered and paramagnetic states 
cannot originate from the response of magnetic domains or domain walls to 
$B_{\rm ext}$.

The presence of such an extremum in $\lambda_Z(B_{\rm ext})$ has not been reported so far. The conventional
Bloch-Wangsness-Redfield theory, which is expected to be valid at least for a temperature sufficiently large
relative to $T_{\rm lr}$, as at 3.2 K, predicts a monotonous decrease of $\lambda_Z(B_{\rm ext})$
as $B_{\rm ext}$ is increased \cite{Slichter96}. In addition, even if the low field part of the data is 
discarded, such a description breaks down because preliminary measurements, see Fig.~\ref{muon_lf}, 
indicates $\lambda_Z(B_{\rm ext})$ to shoot up at large fields, that is, above 1.0 T at 2.4 K. Note that this effect is not due to a change in $\alpha$ 
since $\alpha\approx 2/3$ for all fields from 0.4 T
upwards.

We shall now provide a discussion for the $\lambda_Z$ and $C_p$ behaviors.
As usual for geometrically frustrated magnetic materials, $\lambda_Z$ is
only weakly temperature dependent in zero-field at low temperature. 
This is accounted for by a Raman scattering process involving two magnetic 
excitations with a density of magnetic states characterized 
by an upturn at low energy and a small gap $\Delta$ proportional to the 
temperature \cite{Yaouanc05}, {\sl i.e.} $\Delta = a k_{\rm B}T$. 
A simultaneous fit of both $\lambda_Z(T)$ and $C_p(T)$ in the 
ordered state can be obtained using the density of states $g_m(\epsilon) = 
b_\mu \epsilon^{-1/2} + b_{\rm sh}\epsilon^{5/2}$ for $\epsilon > \Delta$, 
where the first term gives
rise to a temperature independent $\lambda_Z$ and the second term has been 
chosen to reproduce our experimental $T^{7/2}$ dependence in  
$C_p(T)$. In
Figs.~\ref{Sh_neutron}a and \ref{muon_zf}b are shown the results of this
model with $a$ = 0.02 and $b_\mu$ and $b_{\rm sh}$ respectively equal to
0.057 meV$^{-1/2}$ and 73~meV$^{-7/2}$ per Tb atom. 
In addition we take the spectroscopic factor $g$ = 2 and the anisotropy 
and exchange fields equal to $B_e$ = 10 T and 
$B_a$ = 5 T, respectively (see Eq. 1 of Ref. \onlinecite{Yaouanc05}). 
In fact, since, with the assumption $a\lesssim 1 $, 
$\lambda_Z \propto 
b^2_\mu/a^2$, any set of $a$ and $b_\mu$ parameters satisfying 
$b_\mu/a$ $\simeq$ 
3 meV$^{-1/2}$ will equally fit the data. The specific heat 
imposes a
constraint: if $b_\mu$ is too large the first term in the density of states
gives a contribution to $C_p(T)$. 
Therefore our combined set of data enforces 
$b_\mu \lesssim $ 0.06 meV$^{-1/2}$ and hence $a \lesssim 0.02$.

The increase of $\lambda_Z$ at low fields is a definite signature of a field induced increase of the density of excitations at low energy.  
Because of the relatively large Zeeman energy on the system compared to $\Delta$,
the effect of the closure of the gap by the field is negligible. 
Indeed, at $B_{\rm ext} = 10$ mT the Zeeman energy is $3.1 \, \mu {\rm eV}$ which is 
large in comparison to the computed value $< 1 \, \mu {\rm eV}$ for the gap at 0.17 K. 
Experimentally, attributing the stronger relaxation at low field to an increase of $b_\mu$ and
using the value of the slope from Fig.~\ref{muon_lf}, 
for $T< T_{\rm lr}$ we estimate the relative change of $b_\mu$ between 0 and 10 mT to be 0.21.

In conclusion, the magnetic order with a limited correlation length observed in
Tb$_2$Sn$_2$O$_7$ is dynamical in nature and characterized 
by a time $\sim 10^{-10}$ s, much less than previously inferred \cite{Mirebeau05}
but in agreement with our analysis of the neutron results.  This is about an 
order of magnitude longer than estimated for the analogous compound 
Tb$_2$Ti$_2$O$_7$ which does not order \cite{Gardner99a,comments_Tb1}.
Our result implies that the magnetic scattering of the neutrons 
is not purely elastic. Such a dynamical ground state is believed to be at work
in the heavy fermion superconductor UPt$_3$ \cite{Joynt02}. 
A sharp change in the dynamics at the temperature where a strong specific heat anomaly
is detected has already been reported for the pyrochlore compound Yb$_2$Ti$_2$O$_7$ which in contrast displays no magnetic reflections \cite{Hodges02a}. 
Tb$_2$Sn$_2$O$_7$ is a new case since both a specific heat peak and broadened magnetic reflections are detected
at low temperature. Given the time range of the fluctuations, the moments are not as correlated
as it was inferred previously \cite{Mirebeau05}.
Neutron spin-echo experiments may provide
an independent signature of the dynamical nature of the ground state of Tb$_2$Sn$_2$O$_7$.
This dynamics is related to the existence of an appreciable density of magnetic excitations at 
low energy. We have discovered that it can be increased by applying a small magnetic field.

We are grateful to P. Bonville and S. Pouget for useful discussions and S. Sosin for 
complementary specific heat measurements.
P.C.M. Gubbens thanks the Dutch Scientific Organisation (NWO) for its financial support for 
the use of ISIS. This research project has been partly supported by the European Commission 
under the $6^{\rm th}$ Framework Programme through the Key Action: Strengthening the European 
Research Area, Research Infrastructures; contract no: RII3-CT-20030505925.

\bibliography{Tb2Sn2O7_letter}

\begin{thebibliography}{27}
\bibitem[{com({\natexlab{a}})}]{comments_Tb}
\bibinfo{note}{A slight deviation from the stretched-like behaviour has been
  detected at 0.17 K, with the maximum effect for the 70 mT spectrum. It
  manifests itself as a small shoulder between $\sim 0.1-0.3\, \mu{\rm s}$. We
  believe it could reflect the effect of a residual parasitic phase which can
  be hardly detected with powder X-ray scattering.}
\end{thebibliography}

\end{document}